# Magnetic field expulsion in optically driven YBa$_2$Cu$_3$O$_{6.48}$

S. Fava[1,*], G. De Vecchi[1,*], G. Jotzu[1,*], M. Buzzi[1,*], T. Gebert[1], Y. Liu[2], B. Keimer[2], A. Cavalleri[1,3]

[1] Max Planck Institute for the Structure and Dynamics of Matter, 22761 Hamburg, Germany

[2] Max Planck Institute for Solid State Research, 70569 Stuttgart, Germany

[3] Department of Physics, Clarendon Laboratory, University of Oxford, Oxford OX1 3PU, United Kingdom

e-mail: andrea.cavalleri@mpsd.mpg.de, gregor.jotzu@mpsd.mpg.de, michele.buzzi@mpsd.mpg.de

**Coherent optical driving in quantum solids is emerging as a new research frontier, with many demonstrations of exotic non-equilibrium quantum phases. These are based on engineered band structures[1–4], and on stimulated nonlinear interactions between driven modes[5,6]. Enhanced functionalities like ferroelectricity[7,8], magnetism[9–12] and superconductivity[13–16] have been reported in these non-equilibrium settings. In high-T$_c$ cuprates, coherent driving of certain phonon modes induces a transient state with superconducting-like optical properties, observed far above T$_c$ and throughout the pseudogap phase[17–20]. Questions remain not only on the microscopic nature of this phenomenon, but also on the macroscopic properties of these transient states, beyond the documented optical conductivities. Crucially, it is not clear if driven cuprates exhibit Meissner-like diamagnetism. Here, the time-dependent magnetic-field amplitude surrounding a driven YBa$_2$Cu$_3$O$_{6.48}$ sample is probed by measuring Faraday rotation in a GaP layer adjacent to the superconductor. For the same driving conditions that result in superconducting-like optical properties[17–20], an enhancement of magnetic field at the edge of the sample is detected, indicative of induced diamagnetism. The dynamical field expulsion measured after pumping is comparable in size to the one expected in an equilibrium type II superconductor of similar shape and size with a volume susceptibility $\chi_V$ of order -0.3. Crucially, this value is incompatible with a photo-induced increase in mobility without superconductivity. Rather, it underscores the notion of a pseudogap phase in which incipient superconducting correlations are enhanced or synchronized by the optical drive.**

---

[*] These authors contributed equally to this work

A number of recent experiments have made use of ultrashort pulses to dynamically reduce or enhance signatures of superconductivity. For example, irradiation with visible or ultraviolet pulses has been used to study the disruption and recovery of the superconducting state[21-25]. In figure 1(a-c) we summarize the results of one such experiment[21] where a $YBa_2Cu_3O_{6.5}$ ($T_c = 52$ K) thin film was kept at a temperature $T = 10K < T_c$ and photo-excited with near infrared pulses polarized in the *ab*-plane. The dynamics of the optical conductivity was tracked by probing the sample transient optical properties with a time delayed THz pulse. Figure 1b displays the imaginary part of the optical conductivity $\sigma_2(\omega)$ measured before (red line) and after photo-excitation at the peak of the response (blue line). After photo-excitation, the $1/\omega$ divergence of $\sigma_2(\omega)$, characteristic of the superconducting state, was strongly reduced, reflecting the disruption of equilibrium superconductivity. Figure 1c shows the time evolution of $\lim_{\omega \to 0} \omega \sigma_2(\omega)$, a quantity that in equilibrium is proportional to the superfluid density. This measurement displays a prompt reduction of the superfluid density within the first few picoseconds, followed by a slower relaxation back to the superconducting state.

In a series of more recent experiments, mid infrared optical pulses were used to drive $YBa_2Cu_3O_{6+x}$ along the insulating c-axis direction, coupling to apical oxygen vibrations, to coherently modulate the electronic properties. For this type of excitation, heating is limited[26] and new types of coherence can be activated. A transient state with superconducting-like optical properties[17-19] was observed up to temperatures far in excess of equilibrium $T_c$ and throughout the equilibrium pseudogap phase. Representative results are summarized in figure 1(d-f). Single crystals of $YBa_2Cu_3O_{6.48}$ were held at a temperature $T = 100$ K $\simeq 2T_C$ and irradiated with ~500 fs long pulses, polarized along the crystal *c*-axis and centered at 15 μm wavelength, resonant with infrared active modes that modulate the apical oxygen position[27]. The terahertz frequency $\sigma_2(\omega)$ spectra measured before (red line) and after (blue line) photoexcitation are shown in figure 1e. After excitation, $\sigma_2(\omega)$ acquired the same $\sim 1/\omega$ behavior observed along the c-axis in the equilibrium low-temperature superconducting phase[19]. These results, which have been taken as suggestive of photo-induced superconducting correlations, were observed transiently over time windows that range from 1 ps to 5 ps, depending on the duration of the drive[20]. For the experimental conditions explored here, the time-dependent superfluid density (figure 1f) shows that the superconducting-like response is enhanced and reaches a maximum value within ~1 ps, relaxing back to equilibrium on a similar time scale, compatible with the decay of the driven phonon[6].

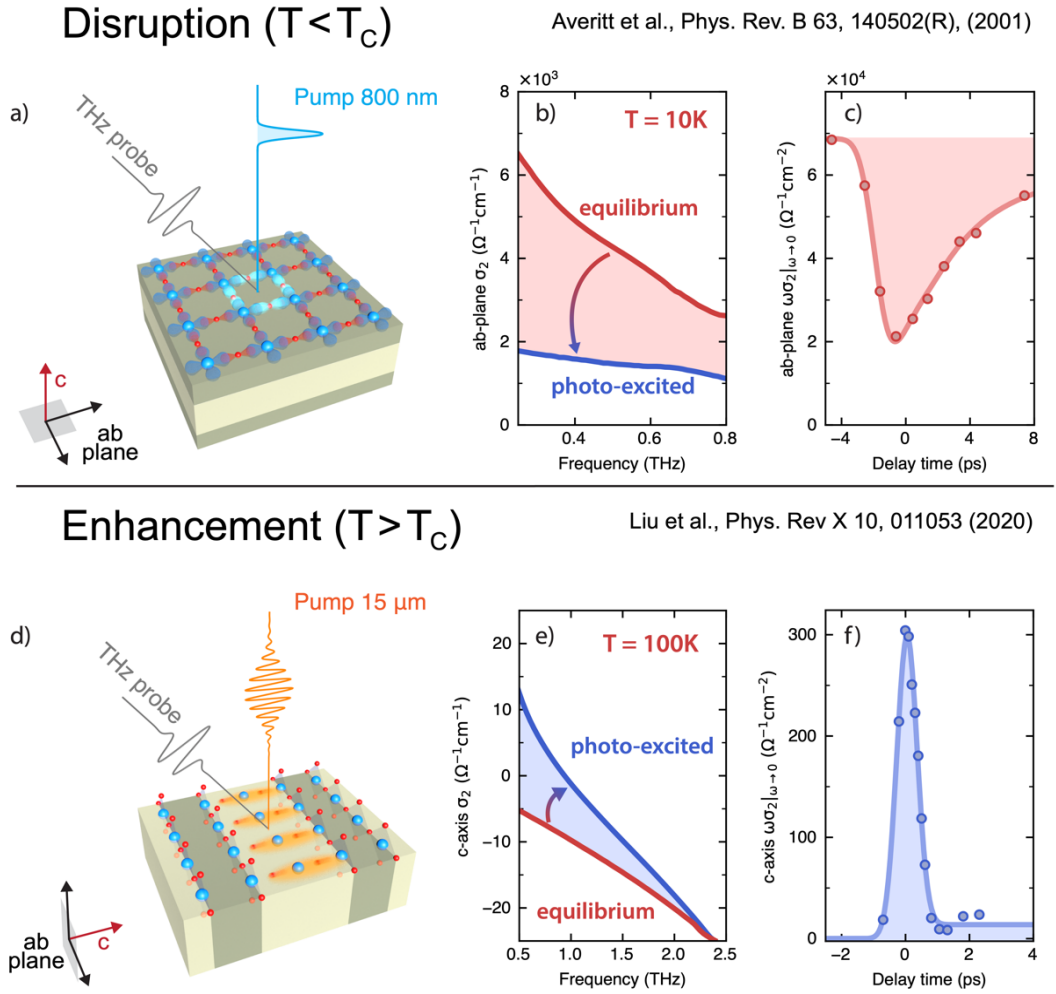

**Figure 1 | Transient optical properties of YBa$_2$Cu$_3$O$_{6+x}$ upon photo-excitation. (a)** Below $T_c$, femtosecond near-infrared pulses polarized in the ab-plane of YBa$_2$Cu$_3$O$_{6.5}$ are used to break superconducting pairs and disrupt superconductivity. The ab-plane optical conductivity of the transient state is probed with THz time domain spectroscopy. **(b)** Imaginary part of the optical conductivity $\sigma_2(\omega)$ measured at equilibrium (red line) and at the peak of the pump-probe response (blue line). At equilibrium $\sigma_2(\omega)$ shows a $1/\omega$ behavior, indicative of dissipationless transport. After photo-excitation, the $1/\omega$ divergence is dramatically reduced. **(c)** Time evolution of $\omega\sigma_2(\omega)|_{\omega\to 0}$, a quantity that at equilibrium is indicative of the superfluid density in a superconductor. The time dependence shows a prompt reduction of the cooper pair density after photo-excitation, persisting for several picoseconds. These data are reproduced from Ref. 21. **(d)** Above $T_c$, YBa$_2$Cu$_3$O$_{6.48}$ is excited with intense mid infrared pulses, resonant with the apical oxygen phonon mode. Broadband THz pulses probe the c-axis optical conductivity of the sample. **(e)** Same quantity as in (b) but measured along the sample c-axis both at equilibrium (red line) and after mid infrared excitation (blue line). **(f)** Same as in (c) but measured along the sample c-axis following mid infrared irradiation. Here, after excitation, a finite transient "superfluid density" appears and persists for $\approx$ 1 ps. The peak value is comparable to that measured at equilibrium below $T_c$ along the c-axis direction in this doping. These data are reproduced from Ref. 19.

In this paper, we explore whether the similarities in optical conductivity between the transient state and the low temperature equilibrium superconductor, extend to the magnetic properties. In analogy with "field-cooled" Meissner diamagnetism, we search for an ultrafast magnetic field expulsion when the material is excited *in a static applied magnetic field*. Under these conditions, non-superconducting high mobility carriers would not modify the magnetic field surrounding the sample, as high conductivity only opposes changes in the magnetic flux. On the other hand, a photo-induced state with superconducting correlations would expel the magnetic field from its interior because of a change in its magnetic susceptibility[28].

To perform these measurements, we adapted existing techniques of magneto-optical imaging[29] and magneto-optic sampling[30], to enable ultrafast magnetic field measurements in a GaP (100) magneto-optic detection crystal. Through the Faraday effect, the polarization rotation of a probe laser pulse yielded the value of the time- and space-dependent magnetic field with sensitivity of better than 1μT.

We first validated the reliability of this technique by measuring the equilibrium superconducting transition in $YBa_2Cu_3O_7$ at equilibrium. The experimental configuration is shown in figure 2a. The sample was a ~150 nm-thick film of $YBa_2Cu_3O_7$ grown on $Al_2O_3$, out of which a 400 μm diameter half disc shape with a well-defined edge was created using optical lithography. The YBa$_2$Cu$_3$O$_7$ film was kept at a temperature $T = 30$ K $< T_c$ with a spatially homogeneous ~2 mT magnetic field applied perpendicular to the plane of the film (vertical direction in the figure) and generated using a Helmholtz coil pair. A ~75 μm thick, (100)-oriented GaP crystal was used as magneto-optic detector and was placed directly on top of the sample. A linearly polarized 800 nm, ultrashort probe pulse was focused to a spot size of ~50 μm on the GaP crystal, impinging at near normal incidence. The Faraday effect induced a polarization rotation on the beam reflected from the second GaP/vacuum interface, providing a measurement of the vertical component of the local magnetic field, averaged over the probed volume, which due to the reflection geometry was traversed twice. In this experiment the polarity of the external magnetic field $B_{ext}$ was periodically cycled, and signals acquired with $+B_{ext}$ applied field were subtracted from those acquired with $-B_{ext}$. In this way, only contributions to the polarization rotation induced by the local magnetic field were measured (see Supplementary Sections S2, S3, and S4 for more details on the experimental setup).

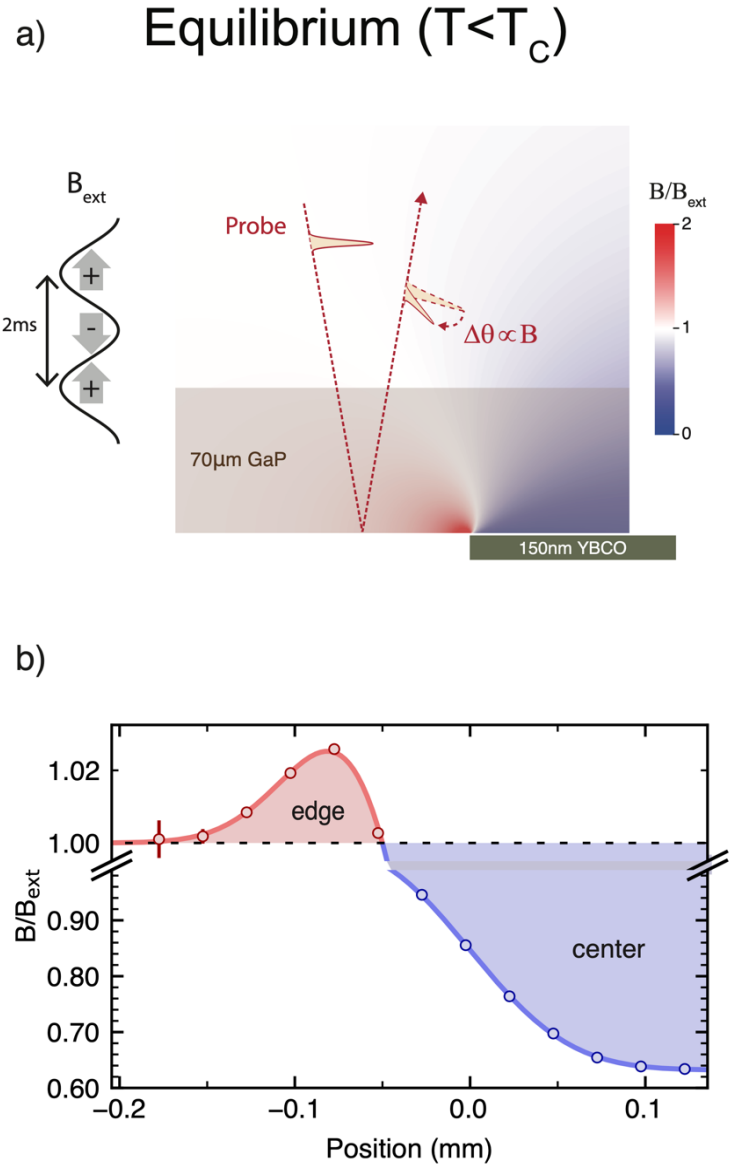

**Figure 2 | Ultrafast Optical Magnetometry probing the magnetic response of a superconductor. (a)** At equilibrium below $T_c$, a lithographically defined YBa$_2$Cu$_3$O$_7$ half-disc of 400 μm diameter expels a static, externally applied magnetic field $B_{\text{ext}}$ (see supplementary S2 for a full 3D view of the geometry). The local changes in the magnetic field are represented using a color plot obtained from a magnetostatic calculation assuming that the superconducting film behaves as a medium with homogeneous magnetic susceptibility $\chi \sim -1$. This spatially inhomogeneous magnetic field is quantified by measuring the polarization rotation, induced by the Faraday effect, of a linearly polarized 800 nm probe pulse, reflected after propagation through a GaP (100) crystal placed in close proximity to the sample. To isolate the magnetic contributions to the polarization rotation, the static applied magnetic field polarity is cycled at a frequency of 450 Hz, while the laser probe pulses impinge on the sample at 900 Hz frequency. **(b)** Ratio of the measured local magnetic field B to the applied one $B_{\text{ext}}$, as function of distance across the edge of the sample. An increased local magnetic field is measured near its edge with vacuum (red) and a reduced one above the sample near its center (blue). The error bars denote the standard error on the mean.

Below $T_c$, the superconductor expels magnetic fields from its interior and changes in the vertical component of the B field can be estimated using a magnetostatic calculation (see supplementary S6 for more details). The results of this calculation are shown in the color plot overlayed to the experimental geometry. The magnetic field outside the sample is expected to be reduced in the center and enhanced near the edge as magnetic flux is expelled from the sample (blue and red regions in figure 2a, respectively). These changes were probed by scanning the probe beam in the horizontal direction and by measuring the sensed magnetic field as a function of distance from the edge.

The results of this measurement are displayed in figure 2b. As predicted, we observed a reduction of the measured magnetic field when measuring above the sample (blue shaded area) and a corresponding enhancement near the edge (red shaded area). The different amplitudes of the effect measured in these two locations are determined by the geometry of the experiment. As shown in the color plot in figure 2a, the decay of the changes in the local magnetic field along the vertical direction is steeper near the edge than above the center of the sample. Because the magneto-optic detector averaged the magnetic field along the vertical direction, a smaller amplitude signal was detected when measuring near the edge where the field experienced a steeper decay (see Supplementary Material section S6).

A second test for time-resolved magnetometry was performed in the conditions of figure 1(a-c). We tracked the dynamics of the magnetic field expulsion when superconductivity was destroyed with an ultraviolet (400 nm) pulse in a $YBa_2Cu_3O_7$ thin film. The geometry of the experiment, shown in figure 3a, was the same as those used in the equilibrium measurements, with the only addition of the pump pulse, which struck the sample from the bottom. Note that the thin $YBa_2Cu_3O_7$ film was completely opaque to 400 nm radiation, and the beam was shaped as a half gaussian to match the half-disc shape defined on the sample. This geometry ensured that the magneto-optic detector never interacted directly with the optical pump (for more experimental details see section S2 and S3 of the Supplementary Material).

The pump induced changes in the local magnetic field were measured as a function of pump-probe time delay in two different positions, near the edge and above the sample. The results of these measurements are displayed in figure 3b. As superconductivity is disrupted (see figure 1(a-c)), the magnetic field penetrates back into the sample within few picoseconds, causing a small decrease in the magnetic field near the edge of the sample (red symbols) and a large increase above (blue symbols). These changes were measured to persist for several picoseconds. Further details about these measurements are described in Supplementary Section S7.

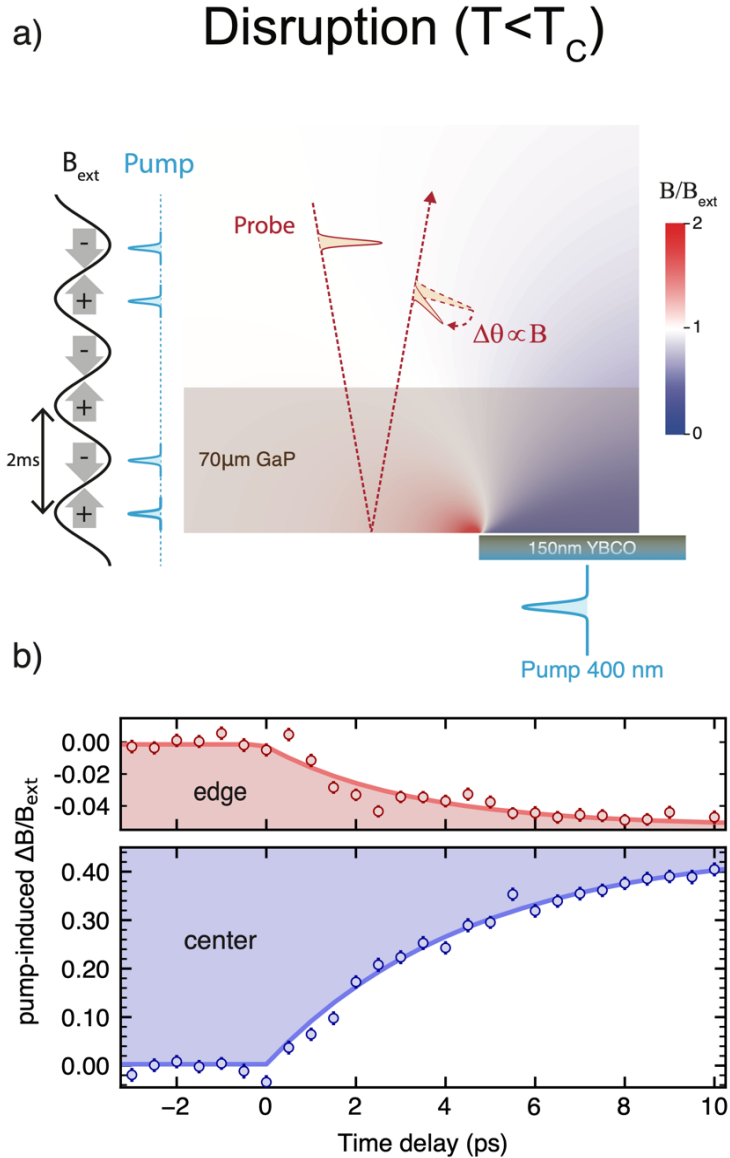

**Figure 3 | (a)** The same YBa$_2$Cu$_3$O$_7$ half-disc shown in figure 2 is cooled below $T_c$ and photo-excited with an ultraviolet laser pulse to disrupt the superconducting state (see supplementary S2 and S7 for a full 3D view of the geometry). The time and spatially dependent pump-induced changes in the local magnetic field are quantified, analogously to figure 2a, through the Faraday effect in a GaP (100) crystal. To isolate the magnetic contributions to the polarization rotation and extract pump induced changes, the applied magnetic field polarity is cycled at 450 Hz frequency while the pump is chopped at 225 Hz frequency. **(b)** Pump-induced changes in the local magnetic field ΔB normalized to the external magnetic field $B_{ext}$, measured near the edge (red) and near the center of the half disc (blue) as function of pump-probe delay. The error bars denote the standard error on the mean.

We next turn to the core observations reported in this paper, and to the measurement of magnetic field expulsion in YBa$_2$Cu$_3$O$_{6.48}$ after excitation with 15 μm mid infrared pulses. For the case of photo-induced superconductivity studied here, we expect to observe pump-induced magnetic field changes in opposite direction compared to the ones shown in figure 3b, that is a positive change near the edge and a negative change above the photo-excited area.

A $YBa_2Cu_3O_{6.48}$ single crystal with an exposed *ac*-surface was held at a series of temperatures $T > T_c$ and irradiated using ~1 ps long pulses centered at ~15 µm wavelength, with a peak field strength of ~2.5 MV/cm. A homogeneous magnetic field $B_{ext}$, tuned to amplitudes between 0 to 12.5 mT, was applied along a direction perpendicular to the sample surface, along the *ab* planes of $YBa_2Cu_3O_{6.48}$. These magnetic field values were lower than the documented[31] $H_{c1}$ for optimally doped $YBa_2Cu_3O_{6+x}$, and comparable to the one measured in this sample.

Local pump-induced changes to this magnetic field were measured using the same ultrafast magnetometry technique discussed above. Figure 4a illustrates the measurement configuration. Due to the thickness of the single crystalline *ac*-oriented sample (~500 µm), being significantly greater than the penetration depth of the mid infrared light (~1 µm), both pump and probe were made to impinge onto the sample/detector from the same side. The detector was placed near the edge of the photo-excited region and was completely shielded from the mid infrared pump by two 30 µm-thick z-cut $Al_2O_3$ crystals, placed above the detector and on its side (see Supplementary Material section S2). Importantly, the $Al_2O_3$ crystals also created a sharp edge for the photo-excited region, a prerequisite to maximize the changes in the magnetic field in its vicinity. Note also that in this experiment the field polarity was periodically cycled, and the pump pulses reached the sample at half the repetition rate of the probe pulses, yielding double-differential pump-probe measurements. This ensured that only effects resulting from a combination of mid infrared excitation and the applied magnetic field were detected. Furthermore, the cut of the GaP crystal was chosen to have zero electro-optic coefficient, to avoid confounding contributions to the probe polarization rotation (see Supplementary Materials S5 for more details).

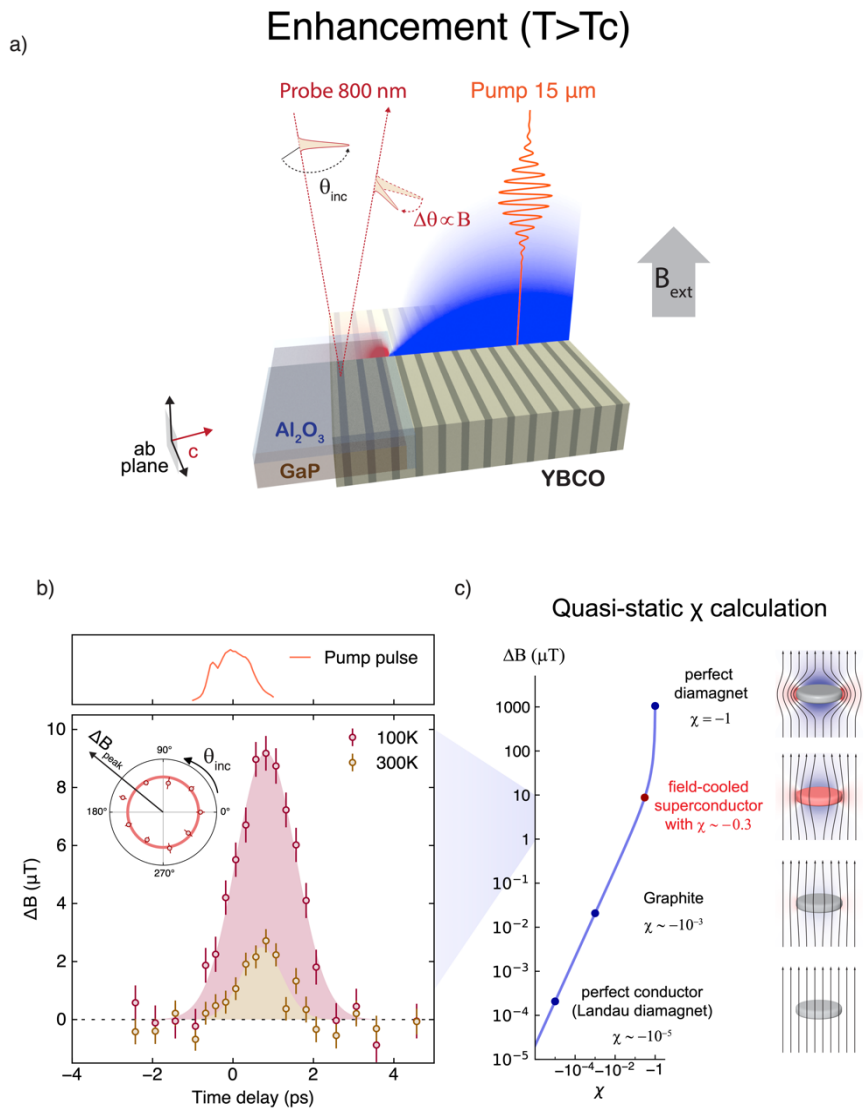

**Figure 4 | Magnetic field expulsion after phonon excitation in YBa$_2$Cu$_3$O$_{6.48}$. (a)** Schematic of the experiment. A thin Al$_2$O$_3$ crystal is placed on top and next to the exposed side of the GaP (100) detection crystal to completely reflect the 15 μm pump and prevent it from generating a spurious non-linear optical response in the GaP (100) crystal. The thin Al$_2$O$_3$ crystal also creates a well-defined edge in the mid infrared pump beam, shaping the photo-excited region into a half disc of ∼ 375 μm diameter. The time dependent changes in the local magnetic field expulsion are probed positioning the probe beam in the vicinity of the edge of the photo-excited region. **(b)** Pump-induced change in the measured magnetic field (Δ$B$) as function of pump-probe delay measured at two different temperatures of 100 K (red) and 300 K (yellow). The upper plot shows the cross-correlation of the pump and the probe pulses measured in-situ in a position adjacent to the sample. Its peak defines the time-zero in the delay scan. The inset shows the dependence of the peak value of the pump-induced magnetic field expulsion (Δ$B_{\text{peak}}$), measured via polarization rotation, on the input polarization angle $\theta_{\text{inc}}$ chosen for the probe pulse. **(c)** Results of a magnetostatic calculation accounting for the geometry and placement of the detector that relate the sensed magnetic field change to a change in the magnetic susceptibility of the photo-excited region in YBa$_2$Cu$_3$O$_{6.48}$. Details of this calculation are contained in Supplementary Section S6. The error bars denote the standard error on the mean.

Figure 4b displays pump induced changes in the local magnetic field measured ~50 μm from the edge created by the MIR mask. The measurements were performed as a function of pump-probe time delay at two different base temperatures of 100 K (red symbols) and 300 K (orange symbols). Upon photo-excitation, a prompt increase of the magnetic field was observed, peaking at a value of ~ 10 μT ($B_{app}$/1000) at 100 K and of ~ 3 μT ($B_{app}$/3000) at 300 K. This transient magnetic field expulsion persisted for ~1 ps, a duration comparable to the lifetime of the superconducting-like conductivity spectra of figure 1f.

A first evaluation of these data is consistent with a photo-induced Meissner-like response. For the experimental geometry of figure 4a we first compare the measured changes in magnetic field to those expected for a *quasi-static* change in magnetic susceptibility. As will be discussed below, this assumption is not entirely justified, although it provides a good estimate to gauge the size of the effect. Assuming that the changes in the photo-excited region are homogenous, and in the fictitious situation of a slow change in the susceptibility of the material, the raw magnetic field changes yield a static induced $\chi_{calc}$ of order -0.3, as shown in figure 4c. We emphasize that this value, is many orders of magnitude greater than what would be observed if the material were made a perfect conductor (assuming weak Landau diamagnetism). Even if it were transformed into a non-superconducting material with the strongest known metallic diamagnetism, such as that observed in graphite, the effect would be at least two orders of magnitude smaller than what observed in this measurement. The colossal photo-induced diamagnetism observed is rather reminiscent of the field-cooled susceptibility of an equilibrium type-II superconductor (see Supplementary Section S1 and S6).

As an additional check, we measured the Faraday rotation induced by the magnetic field expulsion as a function of the incoming probe-polarization angle $\theta_{inc}$. The polar plot in the inset of figure 4b shows the dependence on $\theta_{\text{inc}}$ of the measured magnetic field expulsion at the peak of the response. No variations were observed as $\theta_{\text{inc}}$ was swept from 0 to $2\pi$, corroborating that the signal originated from a Faraday effect. Other effects, such as the Pockels effect, which we stress again should be identically zero because of the crystal cut, would also show a different dependence on the input polarization (see Supplementary Section S5 for a detailed discussion and section S9 for other verification measurements). We next discuss the measured signal taking into account that the changes in diamagnetic susceptibility occur dynamically rather than quasi-statically. Because the change in magnetic field inside the material takes place within approximately 1 ps, the photo-excited area of the sample hosts a rapid change in magnetic field (dB/dt) which is expected to be a source of a picosecond long propagating electromagnetic pulse (see schematic representation in figure 5a). Figure 5b displays pump-induced changes in the local magnetic field measured at three selected distances of 50 μm, 110 μm, and 170 μm from the edge of the photoexcited region (along the horizontal direction, figure 5a). As the propagation distance was increased, the signal was attenuated and peaked at longer delays, as expected for a propagating electromagnetic wave. As displayed in figure 5c, from these values we extracted a propagation speed of $\sim c/n_g$, where $n_g$ is the group index in GaP at $\sim 1$ THz frequency[32].

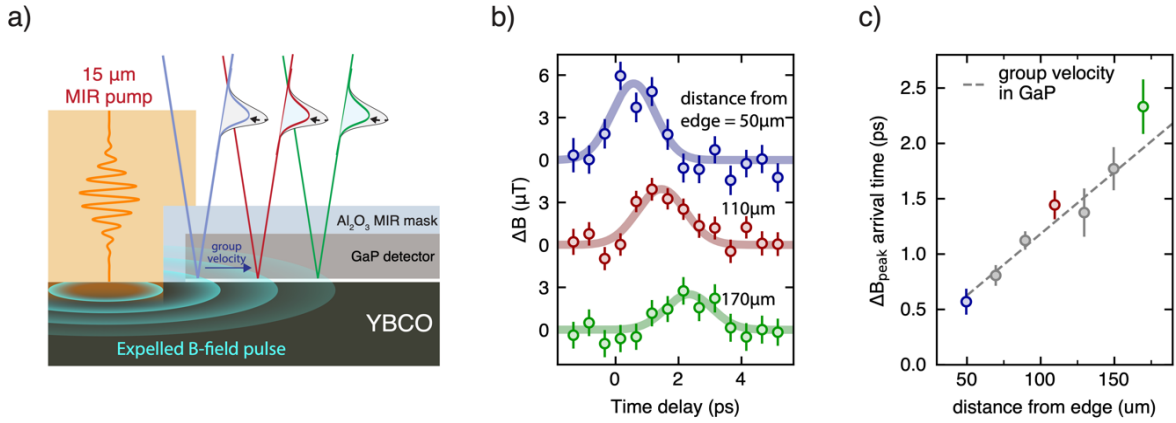

**Figure 5 | Propagating electromagnetic wave emerging from the photo-excited region. (a)** Section view of the experimental configuration. Here, measurements are performed at a base temperature $T = 100$ K, at different distances from the edge of the photo-excited region. **(b)** Pump-induced change of magnetic field as function of pump-probe delay, for three selected positions of 50μm (blue symbols), 110μm (red symbols), and 170μm (green symbols). The solid lines are gaussian fits to the data to extract the peak amplitudes and arrival times. **(c)** Magnetic field peak arrival times extracted from Gaussian fits to the time delay traces measured at different distances from the edge. The grey dashed line shows the expected increase of propagation time with distance based on the group velocity for a 1 THz electromagnetic wave in GaP. The error bars denote the standard error on the mean.

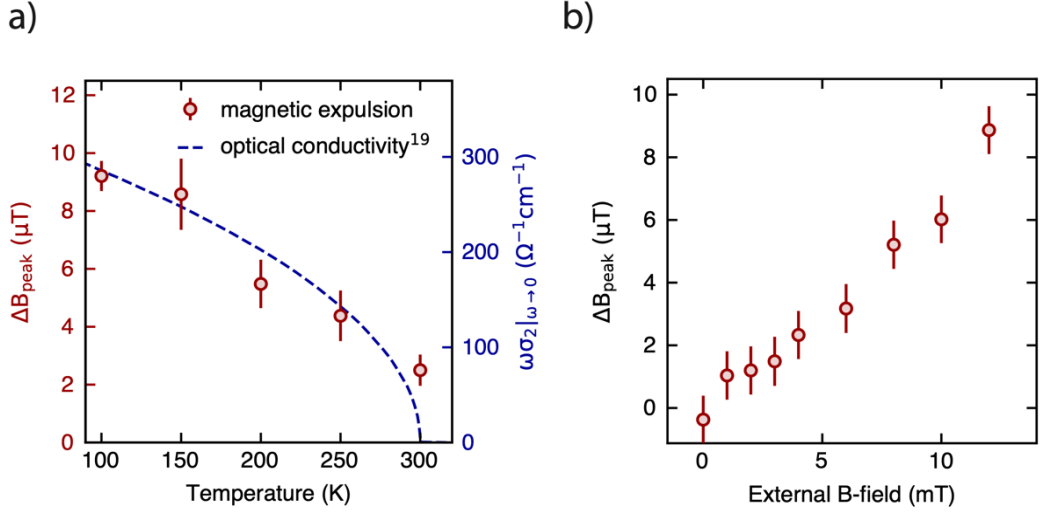

**Figure 6 | Scaling of the magnetic field expulsion with experimental parameters. (a)** Dependence of the photo-induced magnetic field expulsion with temperature (red circles). For each data point a time delay trace was acquired and fitted to extract the peak value $\Delta B_{\text{peak}}$ reported here. See supplementary section S8 for more details. The dashed line shows in comparison the temperature dependence of the photo-induced superfluid density obtained from optical experiments in Ref. 19. **(b)** Peak magnetic field expulsion $\Delta B_{\text{peak}}$ measured for different applied magnetic fields at a fixed time delay $t = 0.75$ ps and at a base temperature $T = 100$ K. The error bars denote the standard error on the mean.

Lastly, we display the dependence of the measured effect as a function of temperature and applied magnetic field. In figure 6a the peak of the pump-induced magnetic field expulsion is displayed as a function of temperature (full time traces are reported in section S8 of the Supplementary Materials). These follow approximately the same temperature dependence of the photo-induced superfluid density, extracted from transient THz data[19], underscoring a common origin for these two physical observations and a correlation with the temperature scale of the pseudogap. As already shown in figure 4, the present data shows how the mid infrared drive applied here can generate a colossal diamagnetic response ($\chi_{\text{calc}}(300\text{ K}) \sim \chi_{\text{calc}}(100\text{ K})/2$) even at room temperature.

Figure 6b shows the same quantity as in figure 6a but measured as a function of the applied magnetic field at a fixed temperature $T = 100$K. Up to the highest B-field that could be generated (~12.5 mT), the value of the expelled field monotonically increases.

We next consider possible explanations for the observed magnetic field enhancement, which is an effect that depends on the applied magnetic field and must be related to a change of magnetic susceptibility in the irradiated area. As discussed above, a quench of the magnetic susceptibility $\chi_v$ from virtually zero to a value of order -0.3, would give rise to a reduction of magnetic field in the photo-irradiated area and an enhancement outside it, explaining the data well. Alternative explanations could involve the interaction of the drive field with small diamagnetic currents that may already exist in the material before excitation. Indeed, if one were to assume that pairing and local superconducting coherences exist throughout the pseudogap phase, it is possible that an amplification mechanism similar to the one discussed for Josephson Plasma polaritons and reported in reference [6] could produce a sizeable magnetic field expulsion qualitatively similar to the one observed.

Both of these effects would underscore some form of photo-induced superconductivity. The first mechanism would be based on a quench of the magnetic susceptibility and hence be compatible with the notion that a true transient superconducting phase is formed, potentially underscoring the existence of a hidden superconductor in the pseudogap phase. The second mechanism would instead amount to an amplification of pre-existing superconducting fluctuations in the pseudogap phase, hence a truly dynamical phenomenon reminiscent of "Floquet" superconductivity, however loosely defined.

Both of these scenarios highlight the highly unconventional nature of this class of physical phenomena, and the role played by coherent electromagnetic fields to engineer quantum materials phases away from equilibrium.

## Acknowledgments

The research leading to these results received funding from the European Research Council under the European Union's Seventh Framework Programme (FP7/2007-2013)/ERC Grant Agreement No. 319286 (QMAC). We acknowledge support from the Deutsche Forschungsgemeinschaft (DFG) via the Cluster of Excellence 'The Hamburg Centre for Ultrafast Imaging' (EXC 1074 – project ID 194651731) and the priority program SFB925 (project ID 170620586). We thank Michael Volkmann, Issam Khayr and Peter Licht for their technical assistance. We are also grateful to Boris Fiedler, Birger Höhling and Toru Matsuyama for their support in the fabrication of the electronic devices used on the measurement setup, to Elena König and Guido Meier for help with sample fabrication, and to Jörg Harms for assistance with graphics. Discussions with Patrick Lee, Eugene Demler, Marios Michael, Dmitri Basov are gratefully acknowledged.

# Magnetic field expulsion in optically driven YBa$_2$Cu$_3$O$_{6.48}$

S. Fava[1,*], G. De Vecchi[1,*], G. Jotzu[1,*], M. Buzzi[1,*], T. Gebert[1], Y. Liu[2],

B. Keimer[2], A. Cavalleri[1,3]

[1] Max Planck Institute for the Structure and Dynamics of Matter, 22761 Hamburg, Germany

[2] Max Planck Institute for Solid State Research, 70569 Stuttgart, Germany

[3] Department of Physics, Clarendon Laboratory, University of Oxford, Oxford OX1 3PU, United Kingdom

e-mail: andrea.cavalleri@mpsd.mpg.de, gregor.jotzu@mpsd.mpg.de, michele.buzzi@mpsd.mpg.de

# Supplementary Material

---

[*] These authors contributed equally to this work

## S1. Sample Growth and Characterization

The optimally doped $YBa_2Cu_3O_7$ thin films were obtained through a commercial supplier (Ceraco GmbH) and grown on R-cut $Al_2O_3$ substrates. The films had a thickness of approximately 150 nm, a sharp superconducting transition temperature at around 85 K and critical current density greater than 2 MA/cm².

The $YBa_2Cu_3O_{6.48}$ single crystals had typical dimensions of ~2 x 2 x 0.5 mm³ and were grown in yittrium-stabilized zirconium crucibles. The hole doping of the Cu-O planes was adjusted by controlling the oxygen content of the CuO chain layer by annealing in flowing $O_2$ and subsequent rapid quenching. A superconducting transition at $T_C$ = 55 K was determined by SQuID DC magnetization measurements, as shown in Figure S1.

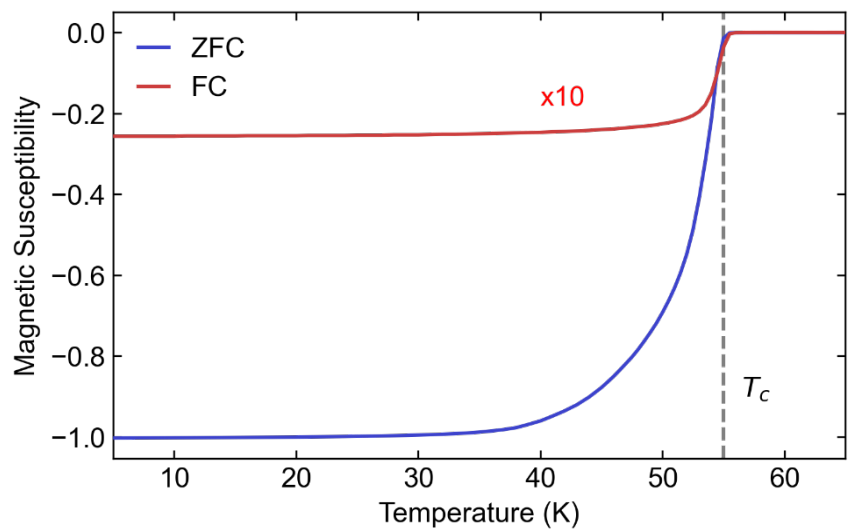

**Figure S1.** Temperature dependent DC magnetization measurements (ZFC: zero field cooled, FC: field cooled) highlighting the superconducting transition in $YBa_2Cu_3O_{6.48}$. The measurements were performed in a 1mT applied field perpendicular to the crystal c-axis.

## S2. Sample Preparation and Experimental Geometries

The $YBa_2Cu_3O_7$ thin films grown on two side polished $Al_2O_3$ were patterned into half-disc shapes using a laser lithography process based on a AZ1512 photoresist mask. After exposure and liftoff, the samples were wet etched using a 1% solution of $H_3PO_4$ acid. After etching, the residual photoresist was removed using acetone and isopropanol. Figure

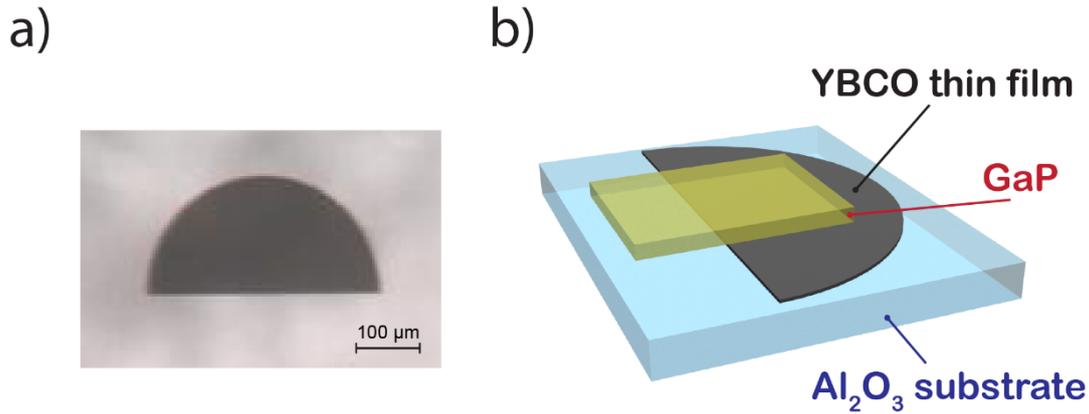

**Figure S2.1. (a)** Micrograph of the YBa$_2$Cu$_3$O$_7$ thin film patterned into a 400 μm diameter half disc shape. **(b)** Sketch of the sample configuration highlighting the positioning of the GaP (100) magneto-optic detector with respect to the YBa$_2$Cu$_3$O$_7$ half disc.

S2.1a shows a micrograph of the YBa$_2$Cu$_3$O$_7$ film after patterning. The thin films and GaP (100) detector were then mounted into an Al$_2$O$_3$ plate that could be fixed directly on the cold finger of the cryostat. The choice of Al$_2$O$_3$ as a material for the sample holder ensures that the effect of eddy currents on the applied field is minimized. A GaP (100) crystal (SurfaceNet GmbH) with a ~1.5° wedge was used as a detector and had a thickness of ~50 μm near the thinner edge. The wedge was used to separate the front and back reflection from each other, therefore allowing to collect only the back reflection containing the accumulated Faraday rotation across the detector. This detector was put in close contact with the sample (see Figure S2.1b), additionally making sure that its back surface and the sample plane were not coplanar, to avoid interference between the reflections from these two surfaces. The gap between detector and the patterned YBa$_2$Cu$_3$O$_7$ film was ~10 μm. This experimental geometry was used for the equilibrium superconductivity and disruption measurements reported in figures 2 and 3 in the main text.

The YBa$_2$Cu$_3$O$_{6.48}$ single crystals were polished after growth to expose an *ac*-oriented surface that allowed access to the crystal c-axis. The single crystal sample was glued on an edge of a half-disc shaped Al$_2$O$_3$ plate. On the top face of the same Al$_2$O$_3$ plate a GaP (100) detector analogous to the one used for the disruption measurements was glued and was in contact with the YBa$_2$Cu$_3$O$_{6.48}$ crystal. A 30 μm thick Al$_2$O$_3$ crystal was placed on top of the GaP crystal and acted as a shield preventing the 15 μm wavelength pump pulses from reaching the GaP detector. A second 30 μm thick Al$_2$O$_3$ crystal was placed on the side to also protect the detector from the side. Note that while transparent for light at 800 nm wavelength $Al_2O_3$ is an almost perfect reflector for 15 μm wavelength pulses and features vanishingly small transmission[1] making it the perfect choice for this application. The two thin $Al_2O_3$ crystals also have the function of shaping the pump beam into a half gaussian with a well-defined edge. A sketch of this experimental geometry is shown in figure S2.2a, alongside with a top view micrograph of the detector and sample assembly in figure S2.2b.

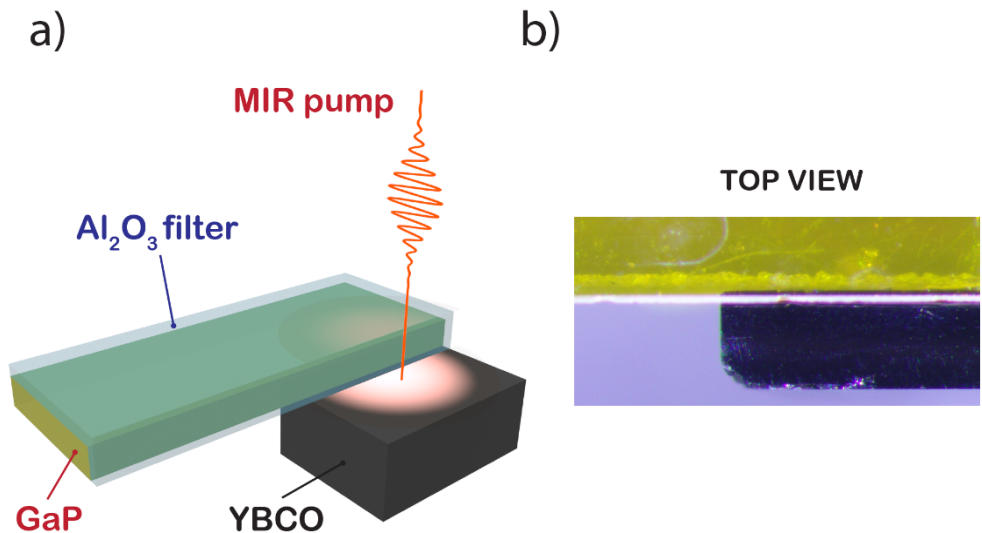

**Figure S2.2. (a)** Sketch of the sample and detector assembly highlighting the positioning of the $Al_2O_3$ filters and GaP (100) detector with respect to the $YBa_2Cu_3O_{6.48}$ single crystal. **(b)** Top-view micrograph of the sample and detector assembly showing the GaP (100) detector (yellow) seen through the $Al_2O_3$ filter and positioned in the vicinity of the $YBa_2Cu_3O_{6.48}$ single crystal (black).

## S3. Experimental Setups and Data Acquisition

The equilibrium spatial scans and superconductivity disruption measurements shown in figure 2 and 3 of the Main Text were performed using the experimental setup sketched in figure S3.1. Ultrashort (100 fs) 800 nm laser pulses were produced starting from a commercial Ti:$Al_2O_3$ oscillator/amplifier chain that produced pulses with energies up to 2 mJ at a repetition rate of 900 Hz. These pulses were split using a beamsplitter into two branches. The lowest intensity branch was used after attenuation for probing the polarization rotation in the GaP (100) magneto-optic detector. To minimize the noise sources in the measurement, the polarization of the beam was set using a nanoparticle high-extinction ratio linear polarizer.

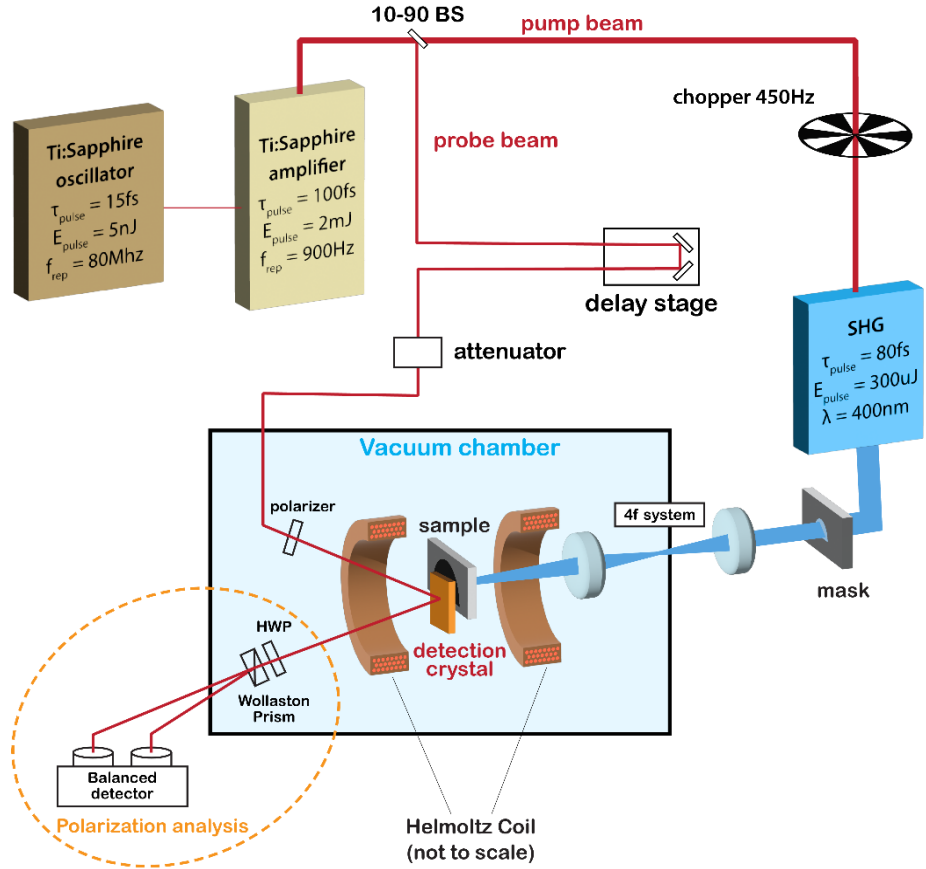

**Figure S3.1.** Experimental setup used for the superconductivity disruption measurements shown in figures 1 and 2 of the main text.

As non-normal incidence reflections introduce a phase delay between *s* and *p* polarization, incidence angle fluctuations can give rise to polarization noise. To minimize this, only reflections close to normal incidence were used in the setup and a commercial system using active feedback was used stabilize the laser beam pointing. After traversing and being reflected from the second surface of the Faraday detector, the polarization state of light was analyzed using a half-waveplate, Wollaston prism and balanced photo-diode setup that allowed us to quantify the Faraday effect in the magneto-optic detection crystal. The higher intensity branch was mechanically chopped at a quarter of the repetition rate (225 Hz) and frequency doubled to obtain 400 nm pulses using a β-$BaB_2O_4$ (BBO) crystal that were used to photo-excite the $YBa_2Cu_3O_7$ thin film samples. A mask, illuminated by these ultraviolet pulses, was imaged onto the back surface of the sample to create a half-gaussian beam with an edge that matched the long edge of the half disc shaped $YBa_2Cu_3O_7$ sample. This, together with $YBa_2Cu_3O_7$ being fully opaque to 400 nm radiation ensured that the GaP detector is not exposed to the pump light. The $YBa_2Cu_3O_7$ thin film samples embedded in the detector assembly (Supplementary Section S2) were mounted on the cold finger of a liquid helium cryostat to allow for temperature control. The cryostat was directly placed in a high vacuum chamber that contained also the part of the optical setup dedicated to polarization analysis to avoid spurious contributions arising from vacuum windows. A pair of coils in a Helmholtz configuration generated a magnetic field at the sample position whose polarity could be reversed at a frequency of 450 Hz. The highest achievable magnetic field was limited by heat dissipation and was ∼3 mT. The sample position was controlled using computer controlled linear translation stages that made it possible to reproducibly move the cryostat and the sample inside the vacuum chamber with ∼10 μm repeatability.

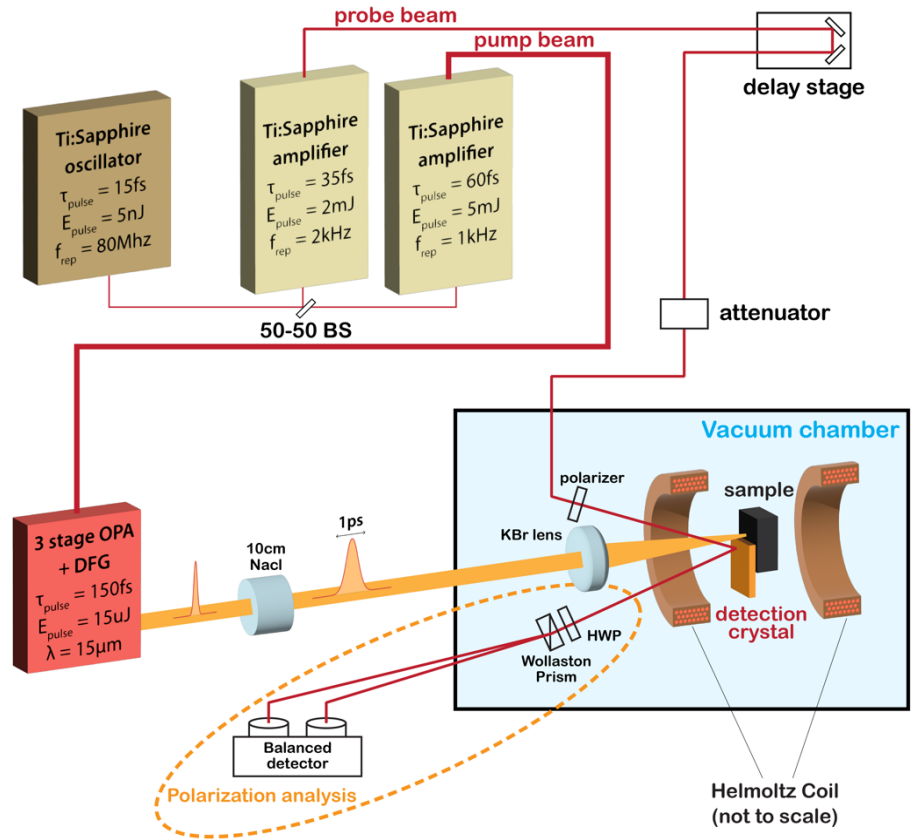

**Figure S3.2.** Experimental setup used for the mid-infrared pump, Faraday effect probe measurements shown in figures 4, 5 and 6 of the main text.

The measurements shown in figures 4, 5, and 6 were carried out with a different experimental setup sketched in figure S3.2. Here, 800 nm pulses were generated using a pair of commercial Ti:Al$_2$O$_3$ amplifiers seeded by the same oscillator to achieve femtosecond synchronization. One amplifier produced 35 fs long, ∼2 mJ pulses at 2 kHz repetition rate and was used for the probe beam. The second amplifier produced ∼60 fs long, ∼5 mJ pulses at 1kHz repetition rate and was used to pump a home built three stage OPA that generated ~2mJ total energy signal and idler pulses. These pulses were mixed in a 0.4mm thick GaSe crystal to obtain ~150 fs long, ~20 μJ energy pulses centered at ~20 THz, close to resonance with the $B_{1u}$ apical oxygen phonon modes of $YBa_2Cu_3O_{6.48}$. These pulses were then chirped using a 10mm NaCl rod to a duration of ~1 ps, in order to match the optimum pulse length for inducing superconducting-like optical properties in $YBa_2Cu_3O_{6.48}$[2]. While the sample stages and cryostat were similar between the two setups, in this case the polarization analysis setup is fully "in-line", i.e. the beam travels directly from the polarizer to the Wollaston analyzer without being reflected by additional mirrors other than the detector. This contributed to further reduce spurious sources of polarization noise. A magnetic field was applied at the sample position using a pair of Helmholtz coils whose polarity was switched at ~10 Hz frequency and could reach a maximum amplitude of 12.5 mT.

In both experimental setups the polarity of the magnetic field is cycled periodically at a sub-harmonic of the pump and probe repetition rates. To obtain differential pump-probe measurements the electrical pulses from the balanced photodetector were digitized using a commercial 8 channel 40MS/s data acquisition card, triggered at the lowest frequency used in the experiment. These signals, acquired in the time-domain, were then integrated, after applying boxcar functions, yielding the amplitude of the signal from the sum and difference channels of the balanced photodetector for each probe laser pulse. Since the acquisition of a full pulse sequence required the acquisition of many pump-probe cycles, the sample clock signal of the data acquisition card is derived using direct digital synthesis from the oscillator repetition rate. In this way drifts in the cavity length and repetition rates of the system do not affect the relative timing of the boxcar functions with respect to the arrival time of the electrical pulse.

## S4. Data Reduction and Analysis

As mentioned in the previous section for all the experiments the polarity of the magnetic field was cycled periodically and measurements with pump and without pump were acquired to yield differential pump probe measurements and isolate contribution to the polarization rotation that were induced by the applied magnetic field. In the following we discuss this approach in detail and the impact it has on the measured quantities.

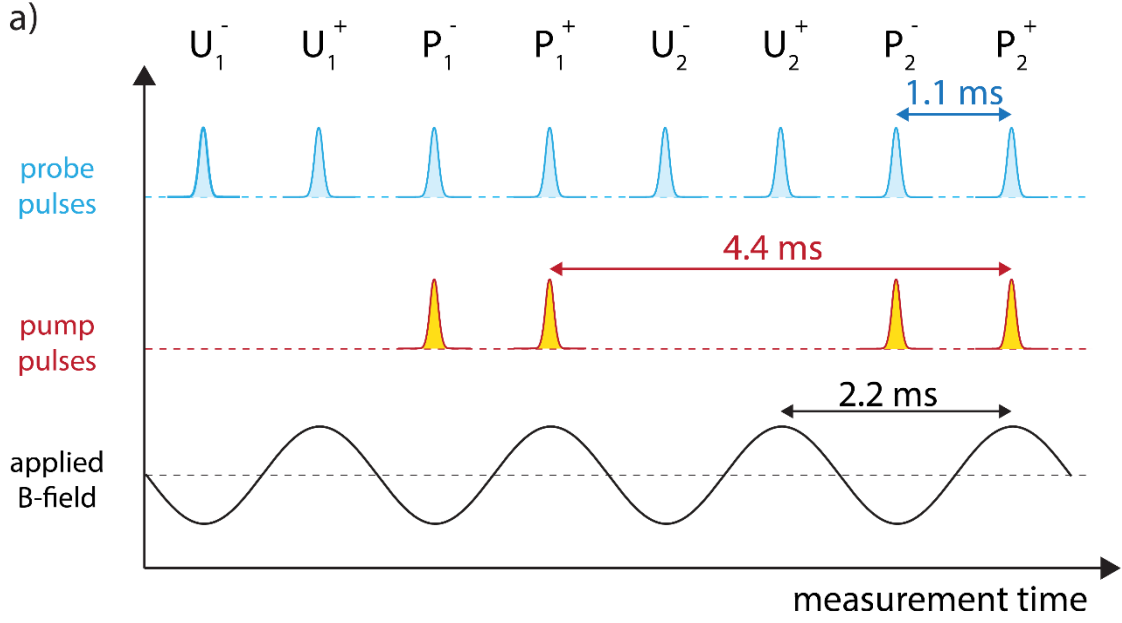

**Figure S4.1.** Timing diagram of the acquisition scheme used for the superconductivity disruption measurements shown in figure 2 and 3 of the main text.

For the measurements shown in figure 2 and 3 of the Main Text the magnetic field polarity was cycled following a sinewave at 450 Hz frequency and the pump was mechanically chopped at 225 Hz. A timing diagram of the acquisition scheme is shown in figure S4.1. The amplitude of the signal of the balanced photodetector difference channel is normalized by that of the sum channel in a pulse-by-pulse manner. For convenience we label these signals as $U_n^\pm$ to indicate those acquired with the pump off for positive and negative polarities of the applied magnetic field and $P_n^\pm$ to indicate the same signals acquired with the pump on. The subscript *n* runs on the pulse number in the sequence. The following quantities are calculated as follows:

$$\Delta\vartheta_{\text{pump-off, n}} = U_n^+ - U_n^- \qquad \Delta\vartheta_{\text{pump-on, n}} = P_n^+ - P_n^- \qquad \Delta\vartheta_{\text{pp,n}} = \Delta\vartheta_{\text{pump-on, n}} - \Delta\vartheta_{\text{pump-off, n}}$$

where $\Delta\vartheta_{\text{pump-off}}$ and $\Delta\vartheta_{\text{pump-on}}$ (averaged over n-pulses) are the magnetic field induced polarization rotations measured with the pump off and on respectively and $\Delta\vartheta_{\text{pp}}$ is the magnetic field induced change in polarization rotation due to the pump. These quantities yielded the amplitude of the magnetic field and its pump-induces changes, after calibration of the Faraday effect in the GaP (100) detector (see Supplementary Section S5). To cancel out residual drifts, the phase of the magnetic field, as well as that of the pump laser, with respect to the probe laser, are periodically alternated between 0 and $\pi$.

The measurements reported in figure 4, 5, and 6 of the main text were acquired using a slightly different scheme to ensure that the sample was excited in a constant magnetic field. Here the probe repetition rate was 2 kHz and the pump struck the sample every second probe pulse (i.e. at 1 kHz) while the magnetic field polarity was modulated following a square wave at a lower frequency of around ~10 Hz. This ensured that the sample was photoexcited in a constant magnetic field. The same quantities as described above were calculated yielding double-differential pump probe measurements that distilled only the contributions to the polarization rotation arising from pump-induced changes in the magnetic properties of the sample.

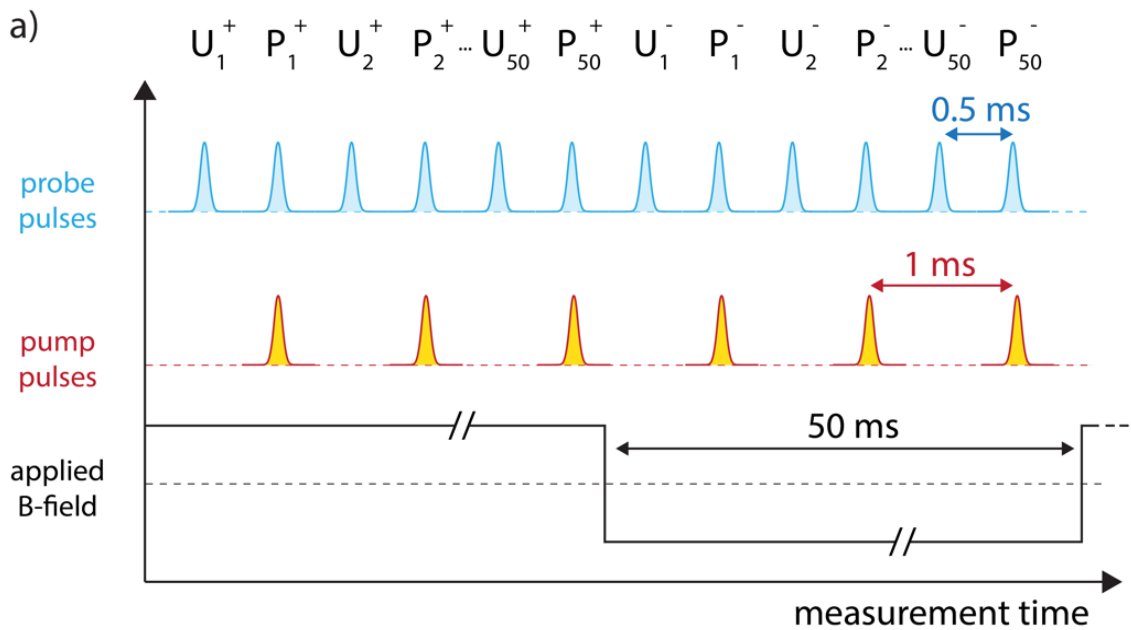

**Figure S4.2.** Timing diagram of the acquisition scheme used for the mid-infrared pump, Faraday effect probe measurements shown in figures 4, 5, and 6 of the main text.

## S5. GaP (100) as a Magneto-Optic Detector

The ultrafast optical magnetometry technique we introduced in the main text relies on the Faraday effect which directly relates the magnetic field applied to a material to the polarization rotation of a linearly polarized beam traversing the medium. This relation is normally reported as:

$$\theta = VBL$$

where $\theta$ represents the rotation of the polarization of the input beam, B is the magnitude of the magnetic field along the light propagation direction inside the medium and L is the thickness of the medium. The proportionality constant V is known as the Verdet constant, which is a material dependent constant depending also on other parameters such as the wavelength of the incoming polarized light.

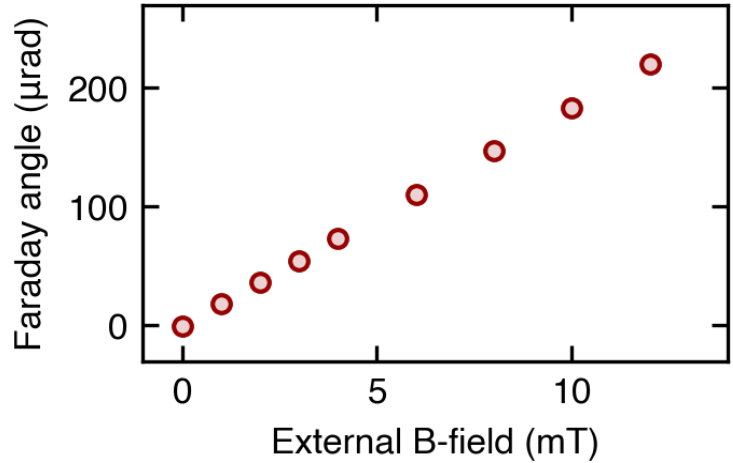

**Figure S5.1.** Sensitivity calibration measurement of the 70 μm thick GaP (100) detectors used for the measurements presented in this work. The measurements were performed at a temperature T = 100K as a function of the applied magnetic field.

In the past the Faraday effect in ferromagnetic crystals and thin films has been used to image the magnetic properties of superconductors at equilibrium[3,4]. While these types of detectors (such as Bi:Y$_3$Fe$_5$O$_{12}$, EuS, and EuSe) offer very high sensitivity ($V \sim 10^5$ rad·T$^{-1}$·m$^{-1}$) they have limited time resolution, down to 100 ps at best, due to the presence of low lying magnetic excitations (e.g. ferromagnetic resonance) at sub-THz frequencies. Diamagnetic II-VI and III-V semiconductors such as ZnSe, ZnTe and GaP have a magneto-optic response featuring Verdet constants that are two to three orders of magnitude smaller than those observed in ferromagnetic materials. Although less sensitive, these materials have the advantage of not being magnetically ordered and offer significantly better time resolution[5,6]. Furthermore, their Verdet constant is mostly temperature independent ensuring a flat detector response in a broad temperature range.

The measurements shown throughout the manuscript were performed using GaP detectors prepared as detailed in Supplementary Section S2. The sensitivity of these detectors was calibrated using the same polarization analysis setup used for the measurements. This was done recording the field-induced polarization rotation applying known magnetic fields with the in-situ Helmholtz coil pair which was independently calibrated using a Lakeshore 425 gaussmeter. The results of this calibration measurement, performed at 100 K is reported in figure S5.1 and accounting for the detector thickness yielded a Verdet constant of ~120 rad·T$^{-1}$·m$^{-1}$, in agreement with reported literature values[7].

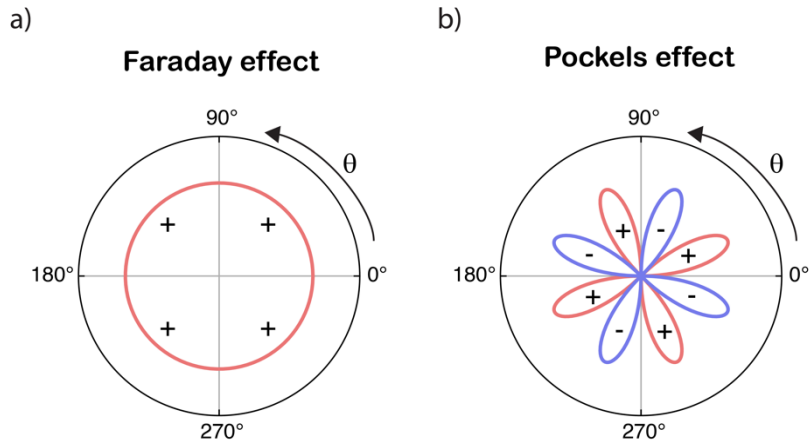

**Figure S5.2. (a)** Input polarization dependence of the measured signal expected for the Faraday effect in a GaP (100) crystal under the presence of a magnetic field parallel to the [100] direction. **(b)** Same as in (a) but due to the electro-optic effect, assuming a finite incidence angle and an electric field applied in the (100) plane.

Since GaP is an optically isotropic material, the Faraday effect is also expected to be isotropic and its strength only depends on the orientation of the light propagation direction with respect to the magnetic field inside the crystal. Hence, the measured magnetic field-induced polarization rotation is expected to be independent on the input polarization angle with respect to the crystal axes as shown in figure S5.2a.

In GaP, inversion symmetry is broken and besides being magneto-optically active it also features an electro-optic (Pockels) effect, that is an electric field induced birefringence. Since ultrafast magnetic field pulses are propagating electromagnetic waves, a time dependent electric field at THz frequencies is also present and care must be taken to distinguish the two effects. In the geometry of our experiment, with an applied magnetic field perpendicular to the plane of the detector, we expect electric fields associated to ultrafast changes in the local magnetic field to be polarized in the plane of the detector. Based on the symmetry of the electro-optic tensor, electric fields polarized in the plane of a (100) oriented GaP crystal will not cause any birefringence for probe beams propagating along the [100] crystal axis.

Small contributions due to misalignment of the crystal orientation, finite incidence angles, or electric fields polarized out of the (100) plane will give rise to a signal at the balanced photo-detector that depends on the angle of the probe beam input polarization, due to symmetry. These contributions can be calculated extracting the field induced birefringence from the eigenvalues and eigenvector of the dielectric impermeability tensor and calculating the expected signal using Jones calculus[8]. For example, in figure S5.2b we report the polarization dependence of the photo-detector signal that is expected to be observed with an electric field in the (100) plane and a finite incidence angle. The signal should exhibit an eight-fold dependence on the input polarization. This is in contrast with the polarization dependence reported in the inset of Figure 4b of the main text and corroborates that the measured signal originated from a Faraday effect.

## S6. Magnetostatic Calculations

The changes in the magnetic field surrounding the sample were calculated in COMSOL using a finite element method to solve Maxwell's equations taking into account the geometry of the experiment. The solution domain was defined as a spherical region of 1mm radius where a constant uniform magnetic field was applied. A half-disc shaped region characterized by a constant, field independent, spatially homogeneous magnetic susceptibility $\chi_v$ was placed in the center of the spherical region and was used to model the magnetic response of either the patterned $YBa_2Cu_3O_7$ thin film or the photo-excited region in $YBa_2Cu_3O_{6.48}$.

While the size of the half disc in the simulation was exactly matched to the one used in the experiments for the patterned $YBa_2Cu_3O_7$, assumptions had to be made regarding the size of the photo-excited region in $YBa_2Cu_3O_{6.48}$. The latter was modelled as a half disc of 375 μm diameter, coinciding with the measured 15 μm pump beam spot size, using different thickness values corresponding to different assumption on the pump penetration depth as discussed below. The weak magnetic response of the substrate or of the unperturbed $YBa_2Cu_3O_{6.48}$ bulk were not included in the modelling as they are expected to be several orders of magnitude smaller due to their much lower magnetic susceptibility. To account for the detector response which, as written in the Main Text, generates a polarization rotation which is proportional to the average of the magnetic field in the volume probed by the light pulse, the results of the calculation were integrated along the detector thickness, yielding from three-dimensional field maps a two-dimensional map of the magnetic field averaged along the detector thickness. These were then convoluted with a two-dimensional Gaussian function to account for the spatial resolution given by the finite size of the focus of the probe beam.

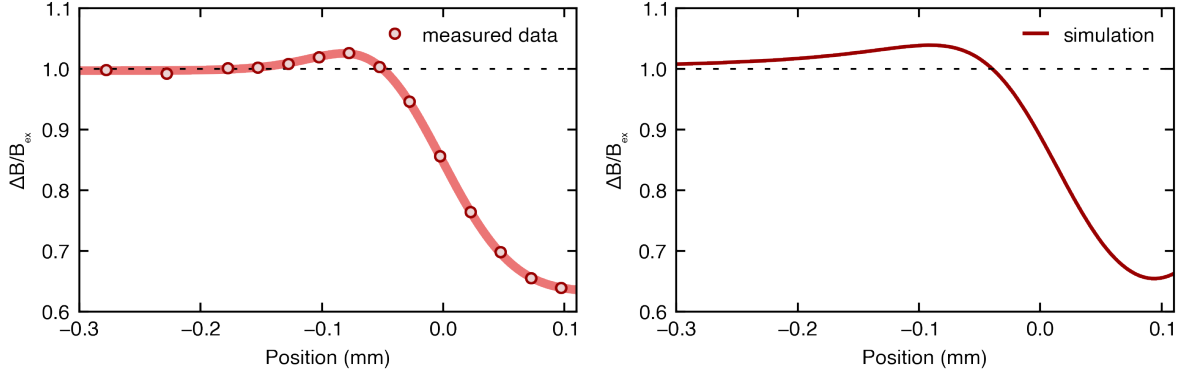

**Figure S6.1. (a)** Ratio of the measured local magnetic field B to the applied one $B_{ext}$, as function of distance across the straight edge of the YBa$_2$Cu$_3$O$_7$ half disc (see figure S2.1). The solid line is a guide to the eye. **(b)** Results of a magnetostatic calculation using the method described above for the set of parameters discussed in the text. Despite the simplicity of the model, the simulated $B/B_{ext}$ shows a good agreement with the experimental data.

Figure S6.1 shows a comparison between a line scan measured across the straight edge of the YBa$_2$Cu$_3$O$_7$ half disc and the results of a magnetostatic calculation performed using geometrical parameters that reflect the experimental conditions. In this simulation $\chi_v$ was varied to achieve the best agreement with the experimental data. We extracted a value for $\chi_v \sim -1$ which is compatible with the zero-field-cooled magnetic properties of YBa$_2$Cu$_3$O$_7$ thin films[9].

We used the same calculations to quantify the magnetic susceptibility that the photo-excited region in YBa$_2$Cu$_3$O$_{6.48}$ should acquire after photo-excitation to produce a magnetic field change equal to the one measured at the peak of the pump-probe response. This was achieved running the calculations for a set of $\chi_v$ values and thicknesses of the photo-excited region to obtain "calibration curves" that related the average magnetic field 50 μm away from the edge to the susceptibility $\chi_v$.

The curve reported in the Figure 4c of the Main Text is calculated under the assumption of a thickness $d$ of the photo-excited region equal to 2 μm, corresponding to the electric field penetration depth of the pump, defined as $d = \frac{c}{\omega \cdot \text{Im}[\tilde{n}_0]}$, where $\tilde{n}_0$ is the stationary complex refractive index of YBa$_2$Cu$_3$O$_{6.48}$ along the c-axis[10] at the pump frequency. This assumption is justified given the sublinear fluence dependence reported in figure S10.1.

In figure S6.2 we test the effect of different assumptions analyzing the dependence of the extracted $\chi_v$ on the thickness of the photo-excited region. Here, we report $B$ vs. $\chi_v$ curves obtained for three different values of:

- $d = 1\ \mu m$, corresponding to the intensity penetration depth of the pump
- $d = 2\ \mu m$, corresponding to the electric field penetration depth of the pump
- $d = 5\ \mu m$, corresponding to the region where ~99% of the pump energy is absorbed.

We stress that independently on the chosen value for the penetration depth, the retrieved value of $\chi_v$ remains in the $10^{-1}$ range, several orders of magnitude higher than the strongest diamagnetic response observed in metallic systems such as graphite (see Figure 4c of the Main Text).

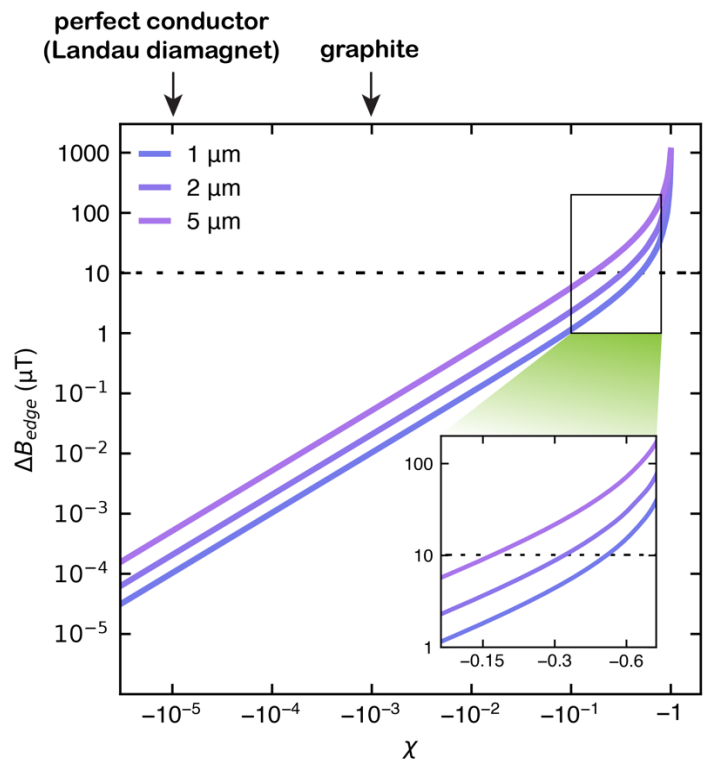

**Figure S6.2.** $B$ vs. $\chi_v$ curves obtained for three different values of $d = 1, 2, 5\ \mu m$. Independently on the assumption chosen, the retrieved value of $\chi_v$ remains in the $10^{-1}$ range, several orders of magnitude higher than the strongest diamagnetic response observed in metallic systems such as graphite (see Figure 4c of the Main Text).

## S7. Spatially Resolved Pump-Probe Scans in YBa$_2$Cu$_3$O$_7$

The data shown in figure 3 were acquired in two different positions (above the sample center and outside of it near the edge) as a function of time delay between the 800 nm probe pulse and the ultraviolet (400 nm) pump pulse. In Figure S7.1 we report spatial dependent measurements of the pump-induced magnetic field changes as superconductivity is disrupted in YBa$_2$Cu$_3$O$_7$ at one selected time delay t = 10 ps. The pump-probe signal shows a spatial dependence similar to that of the static magnetic field expulsion. On the edge of the superconductor, where an enhanced magnetic field is observed at equilibrium, destruction of superconductivity induced a negative pump-induced magnetic field change, indicating that the applied magnetic field penetrated back into the sample. Above the sample center, instead, a reduced magnetic field is observed at equilibrium and disruption of superconductivity induced a positive signal, indicating that magnetic field shielding ceased after photo-excitation. Due to the specific pulse sequence used for the measurement (see Supplementary Section S4) spatially inhomogeneous trapped magnetic flux was present and the amplitude of the magnetic field change is slightly altered compared to what one would expect from the equilibrium measurements.

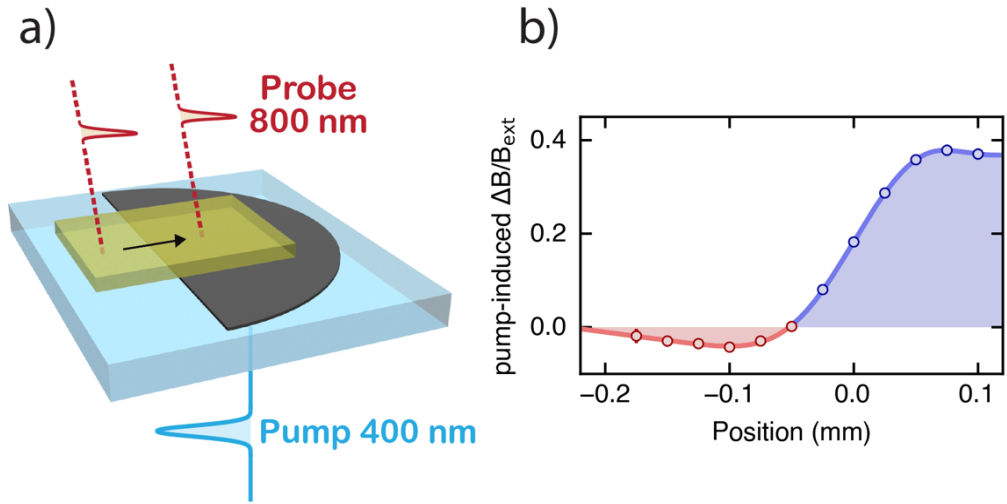

**Figure S7.1. (a)** Sketch of the experimental geometry. The pump beam is kept fixed and the probe beam is moved across the edge of the YBa$_2$Cu$_3$O$_7$ half-disc shaped film. **(b)** Space-dependent pump-induced changes in the local magnetic field measured at T = 30K, with an applied magnetic field of 2 mT and pump fluence of 5 mJ/cm$^2$.

## S8. Temperature Dependent Delay Scans in YBa$_2$Cu$_3$O$_{6.48}$

In analogy with the data in figure 4b, a time delay scan was acquired for each temperature. The peak value of each of these scans was extracted via a Gaussian fit of the data and the peak value was plotted as a function of temperature in Figure 6a. The data reported in figure S8.1 shows that the dynamics of the magnetic field expulsion is mostly independent of temperature and only the peak value reduces as the temperature is increased.

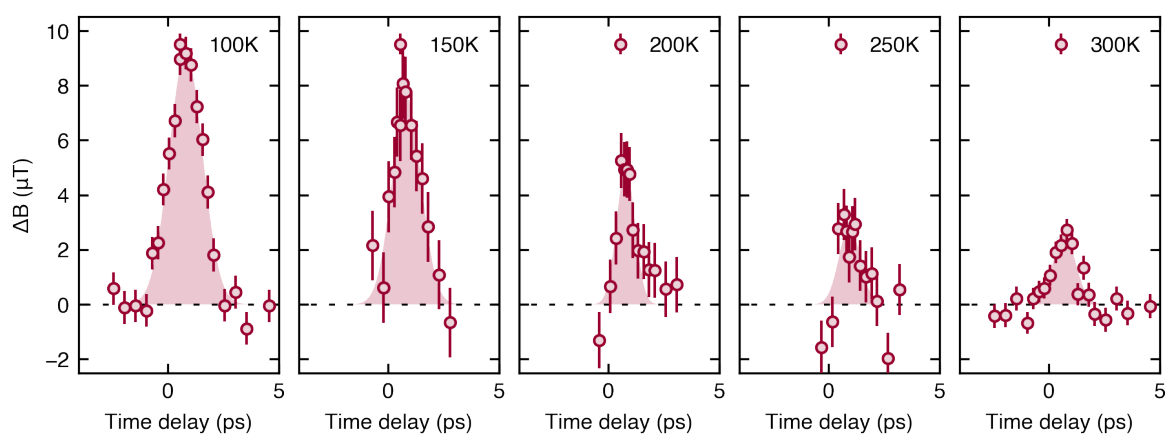

**Figure S9.1.** Time dependence of the local magnetic field near the edge of the YBa$_2$Cu$_3$O$_{6.48}$ crystal measured upon photoexcitation with 15 μm pump pulses at temperatures of 100 K, 150 K, 200 K, 250 K, and 300 K. The shaded areas display the Gaussian fits used to extract the amplitude of the pump-probe signal. These data were measured at a constant pump peak electric field of 2.5 MV/cm, pump pulse duration of ~ 1 ps and an applied magnetic field of 10 mT. The error bars denote the standard error on the mean. The data at 100 K and 300 K have been averaged for significantly longer compared to other temperatures, yielding a noticeably lower noise.

## S9. Additional Spatial Dependences in YBa$_2$Cu$_3$O$_{6.48}$

Figure 5 in the Main Text shows a probe beam spatial dependence performed by moving the probe at progressively longer distances from the edge of the photoexcited region in the YBa$_2$Cu$_3$O$_{6.48}$ crystal. In figure S9 we report additional measurements that were performed at a 50 μm constant distance of the probe with respect to the center of the excitation beam and moving both of them together from a position where the YBa$_2$Cu$_3$O$_{6.48}$ crystal is present underneath the GaP detector layer to one where it is not. A sketch of how this measurement is performed is shown in figure S9.1a and the result of this scan are shown in figure S9.1b. The measurement shows that as the pump and probe are moved on the detector away from the sample the signal vanishes confirming that this signal arises from a magnetic field expulsion in the YBa$_2$Cu$_3$O$_{6.48}$ crystal following photoexcitation, and not from spurious interactions in the magneto-optic detection crystal or in the Al$_2$O$_3$ filter.

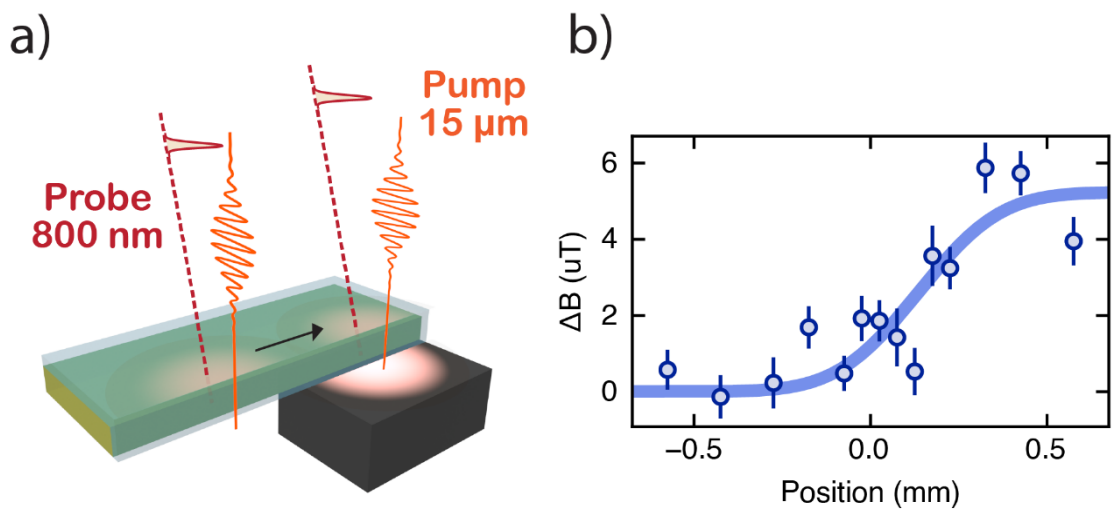

**Figure S9.1. (a)** Sketch of the experimental geometry. The pump and the probe beams are both moved parallel to the edge of the detector from an area where the YBa$_2$Cu$_3$O$_{6.48}$ crystal is present underneath the GaP detector layer to one where it is not. **(b)** Measured pump-induced changes in the local magnetic field at a temperature T = 100 K, time delay t = 0.75 ps, and peak electric field of 2.5 MV/cm. The solid line is a guide to the eye. The error bars denote the standard error on the mean.

## S10. Fluence Dependent Measurements in YBa$_2$Cu$_3$O$_{6.48}$

In figure S10.1 we report the dependence of the measured magnetic field expulsion on the excitation fluence. These data are taken at the peak of the response, at a temperature T = 100 K and with an applied magnetic field of 10 mT, after photoexcitation with 15 μm center wavelength, 1 ps long pulses. The observed scaling appears sublinear in fluence. The data in the figures 4, 5, and 6 was acquired at a fluence of ∼14 mJ/cm², corresponding in this case to a peak electric field of ∼2.5MV/cm.

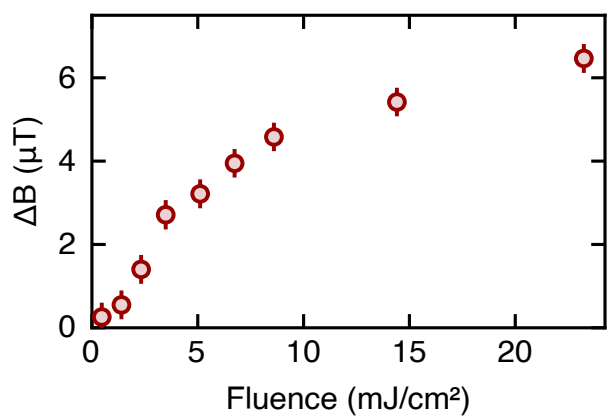

**Figure S10.1.** Fluence dependence of the pump-induced changes in the local magnetic field near the edge of the YBa$_2$Cu$_3$O$_{6.48}$ crystal measured upon photoexcitation with 15 μm pump pulses at a temperature of 100 K. These data were measured at the peak of the response and an applied magnetic field of 10 mT. The error bars denote the standard error on the mean.

# References (Supplementary Information)